\documentstyle[aps,prl,multicol,epsf]{revtex}

\begin{document}
\def\CC{{\rm\kern.24em \vrule width.04em height1.46ex depth-.07ex
\kern-.30em C}}
\def\P{{\rm I\kern-.25em P}}
\def\RR{{\rm
         \vrule width.04em height1.58ex depth-.0ex
         \kern-.04em R}}

\draft
\title{  Entanglement of  Quantum Evolutions}
\author{Paolo Zanardi }

\address{
 Institute for Scientific Interchange  (ISI) Foundation, 
Viale Settimio Severo 65, I-10133 Torino, Italy\\
Istituto Nazionale per la Fisica della Materia (INFM)
}
\date{\today}
\maketitle

\begin{abstract}
The notion of entanglement can be naturally extended from   quantum-states 
to the level of general quantum  evolutions.
This is achieved by considering multi-partite unitary transformations
as elements of a multi-partite Hilbert space and then extended to
general quantum operations. We show some 
connection between this  entanglement
and   the entangling capabilities 
of the quantum evolution.
\end{abstract}
\pacs{PACS numbers: 03.67.Lx, 03.65.Fd}
\begin{multicols}{2}

It is theoretically rewarding to describe  the physical world in terms of subsystems.
It follows that it is  a fundamental kinematical requirement to be able to describe the state-space of a composite system
in terms of the simpler state-spaces associated with  its parts.
In quantum theory a basic axioms states that: 
The state space associated with a bi-partite quantum  system
made out of two subsystems $S_1$ and $S_2$ 
is given by the tensor product of the  state-spaces 
associated with  the $S_i$'s \cite{peres}.

This fact  has been in the recent years
shown to be at the  basis of many of the novel  quantum 
capabilities in information processing and computational power \cite{QC}.
Roughly speaking this is due to the fact that, on the one hand  
the tensor product rule gives rise to exponentially large state-spaces
in which information can be encoded.  On the other hand 
the very existence of {\em entangled} states i.e., not product states
with their characteristic  correlations, amounts 
to a new kind of uniquely quantum computational resource.
A lot of efforts have  been accordingly made aimed to the
understanding of the entanglement of quantum states  \cite{ent}.

Clearly the 
 generation of such entangled states is a  subject that has interest  on its own. 
Therefore more recently   
attention started to be  devoted to the  entangling capabilities
of quantum evolutions. Both from the point of view of average 
entangling power \cite{ZAZA} of a general $d_1\times d_2$ unitary transformation and from the  point view 
of  designing   optimal strategies for entanglement production \cite{cir}. 

In this paper  we address the  related issue of entanglement 
of quantum evolutions.  This notion arises in  a very simple  way
once one  recalls that  unitary operators realizing  the (closed)  quantum
dynamics of a multi-partite system belong   to a multi-partite
state-space as well, the so-called Hilbert-Schmidt space.
One is therefore  naturally led to lift all the notions developed so-far 
for quantum state entanglement to the operatorial realm.
The point is  to see whether, beyond their obvious mathematical meaning,
such concepts at the operator level may provide some novel physical insight.  

In the following we shall move the first steps of   this
programme by introducing an entanglement    measure $E$ over 
the operator space over a bi-partite $d\times d$ state-space.
A quantum protocol that provides   $E$ with a simple operational interpretation  will
be described.
We shall study $E$  analytically 
obtaining explict results for abitrary dimension $d.$ 
Moreover we shall show how this operator entanglement can be extended
to general quantum evolutions; such an extension will allow us to reinterpret 
in a quite  natural way the mapping between quantum operations and quantum states
very recently discussed by Cirac et al \cite{cir1}.    
Finally we shall make a connection between  operator entanglement  
 and the entangling power for bi-partite unitary evolutions introduced in \cite{ZAZA}.

Let us begin by recalling some  basic
definitions  of operator algebras.
Let $\cal H$ be a $d$-dimensional Hilbert space. The algebra  of linear
operators over $\cal H$  has on its own a natural structure of $d^2$-dimensional Hilbert space.
The scalar product between two operators $A$ and $B$ is provided by the Hilbert-Schmidt product:
$<A,\,B>:=\mbox{tr}\,(A^\dagger\,B),\,\|A\|_{HS}:=\sqrt{<A,\,A>}.
$
When  the operator algebra is thought of as endowed with such structure it will be denoted by ${\cal H}_{HS}$
and accordingly the ket notation will be used for operators.        
From the basic fact that the space of linear operators over a tensor product
is given by the tensor product   of linear operators.
it follows 
\begin{equation}
({\cal H}^{\otimes\,2})_{HS}\cong {\cal H}_{HS}^{\otimes\,2}.
\label{basic}
\end{equation}
The space of operators associated with a bi-partite quantum system
is on itself a bi-partite quantum state-space.
From this remark it stems that all the notions and  tools developed so
far for the study of entanglement of quantum states  lift to the operator level in a straightforward way.
In particular one can consider unitaries $U$ in ${\cal H}_{HS}^{\otimes\,2}$
as representing all the possible  evolutions of a (closed) bi-partite quantum system
with state-space ${\cal H}^{\otimes\,2}.$

To address the issue of operator entanglement in a quantitative fashion it is useful
to recall that the Hilbert-Schmidt space  ${\cal H}_{HS}$ is isomorphic to
${\cal H}^{\otimes\,2}$ not just algebraically -- in that they have the same dimension --
but also as Hilbert spaces.
Indeed there  exists  a natural  {\em Hilbert-space} isomorphism $\Psi$ between ${\cal H}_{HS}$
and  ${\cal H}^{\otimes\,2}$ 
given  by 
\begin{equation}
\Psi\colon X\mapsto (X\otimes \openone)\,|\Phi^+\rangle,\quad 
|\Phi^+\rangle:=\sum_{\alpha=1}^d |\alpha\rangle^{\otimes\,2}
\label{map}
\end{equation}
where $\{|\alpha\rangle\}_{\alpha=1}^d$ is an orthonormal basis of ${\cal H}.$
The relation  (\ref{map}) defines indeed
a unitary transformation:
$ \langle\Psi(X),\,\Psi( Y)\rangle = \langle \Phi^+|(X^\dagger\,Y)\otimes \openone
|\Phi^+\rangle=
\sum_{\alpha=1}^d \langle \alpha|\,X^\dagger\,Y\,|\alpha\rangle
 =\, \mbox{tr}\,
(X^\dagger\,Y)=<X,\,Y>.$
Even though  $\Psi$ does not preserve the algebraic structure of ${\cal H}_{HS}$
it has the property of mapping the group ${\cal U}({\cal H})$
onto the manifold of maximally entangled states of ${\cal H}^{\otimes\,2}.$

Moving  to the bi-partite case by tensoring one obtains
a  unitary map $\Psi$ (strictly speaking it would be $\Psi^{\otimes\,2}$)
 between ${\cal H}_{HS}^{\otimes\,2}$ and ${\cal H}^{\otimes\,2}\otimes{\cal H}^{\otimes\,2}\cong{\cal H}^{\otimes\,4}$
that associates with  the  operator
$X$ 
the vector 
\begin{equation}
|\Psi(X)\rangle:= (X_{13}\otimes \openone_{24})\,|\Phi^+\rangle^{\otimes\,2}, \quad (X\in{\cal H}_{HS}^{\otimes\,2})
\label{rewritten}
\end{equation}
where the indexes denotes the factors in ${\cal H}^{\otimes\,4}$
on which the corresponding  operators have a non-trivial action.

The extension of $\Psi$  to general quantum operations  i.e.,completely positive (CP) maps 
is  performed by observing that, in view 
of the operator sum representation theorem  \cite{KRA}, one has
that any CP-map $T$ acting on quantum states $\rho,$ can be written
as  
$T(\rho)=\sum_i  A_i\,\rho\,A_i^\dagger, \,(A_i\in{\cal H}_{HS}).$ 
This allows to associate with $T$ the   operator over  ${\cal H}^{\otimes\,4}$ given by
\begin{equation}
\sum_i  |\Psi(A_i)\rangle\langle\Psi(A_i)|=(T_{13}\otimes\openone_{24})(|\Phi^+\rangle\langle\Phi^+|)^{\otimes\,2}.
\label{gen-op}
\end{equation}
In particular  when  $T$ is {\em pure} i.e., just one $A_i$
one gets a pure state consistently with (\ref{rewritten}).
From  equation (\ref{gen-op}) it is straightforward  to realize  that the above mapping
is -- up to normalization factor --   the one-to-one correspondence between
quantum  operations  and quantum states
discussed  in Ref. \cite{cir1} by Cirac et al for studying separability properties
of  quantum evolutions [see their Eq. (4)].
The idea is to extract information about the CP map $T$
from the (unnormalized) state $\Psi(T)$ taking adavantage from the large
set of tools developed to the date for studying entanglement of quantum states.  
In particular in Ref. \cite{cir1} have been presented  
protocols based on local operations and classical communications
to implement a non-local transformation $U$    
by sharing the entangled state $|\Psi(U)\rangle.$ 
It is remarkable that such a powerful correspondence  between quantum operations 
and positive operators simply stems from  the  natural  extension of the basic map  (\ref{map}).

Given an orthonormal  basis $\{e_j\}_{j=1}^{d^2}$ for ${\cal H}_{HS}$ any 
unitary can be written as $U=\sum_{i,j=1}^{d^2} \lambda_{ij} e_i\otimes e_j.$
Since the mapping $\Psi$  associates with $U$ a state with the same $\lambda_{ij}$'s
i.e.,
$|\Psi(U)\rangle:=\sum_{i,j=1}^{d^2} \lambda_{ij}\, |\Psi(e_i)\rangle \otimes |\Psi(e_j)\rangle,
$
it should be clear that the entanglement properties of $U$ and
$|\Psi(U)\rangle$ are the same.
Indeed all the entanglement measures $E(U)$
depend just on the singular values $\{\lambda_k\}_{k=1}^r$
of the matrix $\lambda=(\lambda_{ij})$ ($r=d^2- \mbox{dim Ker}\,\lambda$ is the rank of $\lambda).$
In particular $U$ and $|\Psi(U)\rangle$ have the same
Schmidt decomposition \cite{peres}.
Once that an entanglement measure $E$ is given
it makes therefore sense to define $E(\Psi(U))$ as the entanglement of the quantum evolution
$U.$ The generalization to an arbitrary  $CP$-map $T$ is obtained by the 
formula $E(T):=E(\Psi(T))$ where now $E$ in the lhs is mixed state-entanglement \cite{ent-meas}.

We observe that from the fact that entanglement measures for states
are not increasing  under local operations and classical communications (LOCC) 	 
it follows that the entanglement of an operation $T$ does not increase
if $T$ is followed by LOCC. This is easily seen as follows.
Let $L=\sum_i A_i\otimes B_i$ be a  LOOC transformation.
From the identity $|\Psi(X Y)\rangle= (X_{13}\otimes\openone_{24})\,|\Psi(Y)\rangle$
it follows  that $|\Psi(L T)\rangle =\tilde L\,|\Psi(T)\rangle,$ where
 $\tilde L:= \sum_i (A_i\otimes \openone)\otimes(B_i\otimes\openone).$
Since this latter map is LOOC for the bi-partite system ${\cal H}^{\otimes\,2}\otimes{\cal H}^{\otimes\,2}$
one has $E(|\Psi(L T)\rangle)=E( \tilde L\,|\Psi(T)\rangle)\le E( |\Psi(T)\rangle).$

In the remainder of the paper we shall adopt as entanglement measure
of the (normalized) $|U\rangle\in {\cal H}_{HS}^{\otimes\,2}$ the linear entropy 
of the reduced density matrix 
\begin{equation}
E(U):= 1-\mbox{Tr}\, \rho_U^2,\quad  \rho_U:=\mbox{Tr}\,|U\rangle\langle U|.
\label{lin-ent-op}
\end{equation} 
Even though the information theoretic content of 
the functional $E$ 
is less direct than
the von Neumann entropy $S(\rho_U)=-\mbox{Tr}\,(\rho_U\ln\rho_U)$
the linear entropy has the distinct advantage of being a {\em polynomial} in $U.$ 
This algebraic simplicity makes possible to find for 
(\ref{lin-ent-op}) explicit expressions from which the main features of  our  operator entanglement measure
can be easily derived. 

Let $T_{13}$ be the permutation (swap) operator between the first and the third factor of
${\cal H}^{\otimes\,4}$ and let us denotes by $\hat T_{13}$ its adjoint action i.e., 
$\hat T_{13}(X):= T_{13}\,X\,T_{13}.$
Reasoning as in Ref. \cite{ZAZA} it is not difficult to show that one can write
\begin{equation}
E(U)=1 -\frac{1}{d^4}< U^{\otimes\,2},\,\hat T_{13} ( U^{\otimes\,2})>.
\label{op-ent} 
\end{equation}
Notice that the term $d^4$ is nothing but the normalization factor of $U^{\otimes\,2}$ 
($\|U\|^2_{HS}=\mbox{tr}\,( U^\dagger\,U)=\mbox{tr}\,\openone=d^2$),
for a generic -not unitary - $X$ it must be replaced by $\|X\|^4.$

Equation (\ref{op-ent})  allows us to give to  our
measure of operator entanglement a direct operational meaning.
To this aim we introduce   the projectors $P_{13}^\pm:= 2^{-1}(\openone\pm T_{13})\otimes\openone_{24}$
over the subspaces corresponding to the eigenvalues $\pm 1$ of $T_3.$
The operator $\rho_{13}^+:=2\,P_{13}^+/[d^3\,(d+1)]$ therefore represents the uniform state over the
eigenvalue $1$ subspace.
Eq. (\ref{op-ent}) can be cast in the form
\begin{equation}
E(U)=  2\,N_d < U^{\otimes\,2}\,\rho_{13}^+\, U^{\dagger \otimes\,2},\,P_{13}^->.
\end{equation}
From this there  follows that  (apart from the numerical factor $N_d:= (d+1)/d$),
$E(U)$ can be viewed - and then in principle measured -  as the  probability of  success of the following
protocol in ${\cal H}^{\otimes\,4}$

{a)} Prepare the state $\rho_{13}^+$;
{ b)} Let it evolve it by $U^{\otimes\,2};$
{c)} Project on the eigenvalue $-1$ eigenspace of $T_{13}.$

Now we shall
derive the properties of $E\colon {\cal H}_{HS}^{\otimes\,2}\mapsto \RR$
directly form Eq. (\ref{op-ent}). 

a) First of all let us  observe that from the 
relation $[T_{13},\,(U_1\otimes U_2)^{\otimes\,2}]=0,$ 
it follows that $,\forall U_1, U_2\in{\cal U}({\cal H}$ 
\begin{equation}
E[(U_1\otimes U_2)\,U]=E[U\,(U_1\otimes U_2)]=E(U).
\label{loc-inv}
\end{equation}
This feature is not a peculiar property
of the linear entropy; it nothing but the invariance of $E$ under the {\em local} unitary transformations of 
${\cal H}_{HS}^{\otimes\,2}.$ 
In other words $E$ is constant along  the orbit  of  unitary elements  generated by the ${\cal U}({\cal H})^4$- action
in ${\cal H}_{HS}^{\otimes\,2}$ given by
$\prod_{i=1}^4 U_i \times U  \mapsto (U_1\otimes U_2)\, U\,(U_3^\dagger\otimes U_4^\dagger).
$
These transformations 
define a $4\,d^2-$dimensional subgroup of ${\cal U}({\cal H}_{HS}^{\otimes\,2})$,
that has the peculiar property of mapping unitaries onto unitaries.
The  orbit of  $U$ will be even referred to as the local equivalence class of $U.$

b) 
From $d^2=\|U\|^2_{HS},\, \| \hat T_{13}\|=1$ and using  Cauchy-Schwarz inequality
one has $< U^{\otimes\,2},\,\hat T_{13} ( U^{\otimes\,2})>\le \|U\|^2_{HS}\,
\| \hat T_{13}(U^{\otimes\,2})\|_{HS}\le  \| \hat T_{13}\|\,\|U\|_{HS}^4=d^4.$
Therefore $E(U)\ge 0.$ Notice that $E$ is also invariant under hermitean conjugation: $E(U)=E(U)^*=E(U^\dagger).$

c)  From the previous point  it follows that 
\begin{equation}
E(U)=0\Leftrightarrow \hat T_{13} ( U^{\otimes\,2})= U^{\otimes\,2}\Leftrightarrow [T_{13},\,U^{\otimes\,2}]=0,
\label{fixed}
\end{equation}
On the other hand point  a) ensures  that one can consider, without loss of generality, just the transformations with the form
(Schmidt decomposition) $U=\sum_{k=1}^r \lambda_k e_k\otimes e_k, \,(r\le d^2).$
Inserting this expression in the fixed-point equation (\ref{fixed})  one finds
$k\neq h\Rightarrow \lambda_k\,\lambda_h=0$ that in turn implies
one must have just one non-vanishing Schmidt coefficient. This means that  $U$ is a tensor product
i.e., the zero locus of $E$ is the local equivalence class of the identity.
Notice that the separability condition (\ref{fixed}) applies to all
 pure operations $T.$

d) Since $E(U)=1-\sum_k |\lambda_k|^4$ and $\sum_k |\lambda_k|^2=1$ one recovers
the  well-known upper bound $E(U)\le  1-1/d^2.$
Such a bound is met by all the elements in the unitary
orbit of the swap operator in ${\cal H}\otimes{\cal H}$.
Indeed from  Eq. (\ref{op-ent}) one has:
$E(S) =1-d^{-4} \mbox{tr}\,[(S\otimes S)\,T_{13}\,(S\otimes S)\,T_{13}]
=1-d^{-4} \mbox{tr}\,[T_{24}\,T_{13}]=1-1/d^2.
$
Obviously these maximally entangled transformations
are the ones having  $d^2$ non-vanishing Schmidt coefficients with the same amplitude. 

e) The manifold  ${\cal U}({\cal H}^{\otimes\,2})$ endowed with the Haar measure $dU$
\cite{CORN} becomes a probability space over which the operator entanglement (\ref{op-ent})
defines  a random variable.
 Resorting to group-theoretic arguments it is possible
to  compute the average of $E$ explicitly \cite{expl}:
\begin{equation}
\overline{E(U)}^U:= \int_{{\cal U}({\cal H}^{\otimes\,2})} d U\, E(U) = \frac{d^2-1}{d^2+1}.
\label{ave}
\end{equation}

It is interesting to notice that unitary operators have on average higher entanglement than generic operators.
Taking the (uniform) average of (\ref{op-ent}) 
over the {\em full} unit ball of  ${\cal H}_{HS}^{\otimes\,2}$ 
one obtains 
$\overline{E(X)}^X=1-<T_{13}|\,Q\,|T_{13}>$ in which   
$Q:= \int_{\|X\|_{HS}=1}
dX |X><X|^{\otimes\,2}.$
Now it is easy to see that
$Q\ge 1/d^4 \pi$  in that the latter operator is just the restriction to the unitary submanifold
of the unit ball in ${\cal H}_{HS}$ of the
same integral of the former.
It follows that
$ <T_{13}\,| Q\,| T_{13}>\ge <T_{13}\,| \pi| T_{13}>,$
which  in turn implies the  announced inequality.

In order to provide some exemplifications of  
 the measure  (\ref{op-ent}) now we consider a couple of very simple cases

1)  Let $\{ \Pi_\alpha\}_{\alpha=1}^r$ be a set of orthogonal projectors such that $ \sum_\alpha \Pi_\alpha=\openone$ 
and $\{U_\alpha\}_{\alpha=1}^d$ is a set of orhogonal unitaries. One can write the controlled unitary operation over 
${\cal H}^{\otimes\,2}$
$U=\sum_\alpha \Pi_\alpha\otimes U_\alpha.$
 one finds that 
$E(U)= 1-1/d^2 \sum_{\alpha} |\mbox{tr}\, \Pi_\alpha|^2.$
Of course  the most (least)  entangled situation i.e., $E(U)=1 -1/d$ ($E=0$), corresponds to having  all the 
 $\Pi_\alpha$'s  one-dimensional ($r=1, \Pi_1=\openone$).

2) Let $\{U_\theta\}_{\theta\in[0, 2\,\pi)}$ be the one-parameter family of  $2\times 2$ unitary transformations
given by 
$U_\theta:=\exp[i\,\theta \, \sigma_z^{\otimes\,2}]=
\cos (\theta) \openone^{\otimes\,2}+i\,\sin(\theta) \sigma_z^{\otimes\,2},
$
where 
$\sigma_z=\mbox{diag}\,(1,\,-1).$
It is straightforward  to obtain:
$E(U_\theta)= 2^{-1} \sin^2 (2\,\theta),$
which clearly displays  the separable (maximally entangled) character
of  the $U_\theta$ for $\theta=0,\,\pi/2$ ($\theta=\pi/4$).

It is worthwhile to mention that the entanglement measure (\ref{op-ent}) has a simple group-theoretic content
in that  it is twice
the expectation value of the projector $P_{13}^-$
on the (normalized) state $|U\rangle^{\otimes\,2}.$
The more this latter state is antisymmetric with respect
to the action of the swap $T_{13}$ the more  it is entangled.
In particular for an unentangled $|U\rangle$  one has that $|U\rangle^{\otimes\,2}$
is completely symmetric and  one gets the quadratic relation  Eq. (\ref{fixed}). 
Interestingly enough  this  characterization of product states
extends to the multi-partite case.
Let ${\cal H}\cong (\CC^d)^{\otimes\,N},$ and $T_{i, i+N}$ be the swap between the $i$-th and the $i+N$-th factor
in ${\cal H}^{\otimes\,2}.$
Since $|\Phi\rangle$ is a product state {\em iff } the $N$ single-subsystem reduced density matrices
are one-dimensional  projectors, one finds  that    for a $|\Phi\rangle$ to be a product a  necessary and sufficient condition is
\begin{equation}
(\openone -T_{i, i+N} )\,|\Phi\rangle^{\otimes\,2} = 0,
\label{pureprodu}
\end{equation}
where $i=1,\ldots,N-1.$
These equatione are just the operator  version of the condition given in Ref. \cite{luc}.

We finally discuss the relation of the  entanglement of  operator $U$
with its {\em entangling power} \cite{ZAZA}.
It must be stressed that such a relation cannot be trivial. 
Indeed the more entangled is an operator  the more it is non-local,
but this does not mean that the greater are its entangling capabilities
[at least in the sense discussed in \cite{ZAZA}].
For example the swap operator $S$ 
maps product states onto product states  and therefore
has not direct entangling capabilities.
On the other hand we have seen that $S$ is maximally entangled.

In Ref. \cite{ZAZA} we defined the entangling power $e_p(U)$ of a unitary $U$
over ${\cal H}^{\otimes\,2}$ 
as the average of the entanglement $E(U\,|\Psi\rangle),$
where the $|\Psi\rangle$'s are product states generated according to 
some given probability distribution $p.$   
By choosing for $E$ the linear entropy  of the reduced density matrix
and using an uniform i.e., $U(d)\times U(d) $-invariant $p$  for the $|\psi\rangle$'s
one finds \cite{ZAZA}
$e_p(U)= 1-d^4\,< U^{\otimes\,2}\,\rho^+_{13}\,\rho^+_{24}\,U^{\dagger\otimes\,2},\,T_{13}> 
.$
Comparing this equation with (\ref{op-ent}) straightforward algebra reveals that
\begin{equation}
e_p(U)= N_d^{-2}\,[E(U)+E(U\,S)-E(S)]
\label{ep}
\end{equation}
The $U$-dependent part of the
entangling power of the evolution $U\in{\cal U}({\cal H}^{\otimes\,2})$
is proportional to the operator entanglement of $U$ 
averaged with respect  to the  multiplicative action of the 
permutation group ${\cal S}_2:=\{\openone,\,S\}.$
Another way to express Eq. (\ref{ep}) is 
as the average of $E$ along the ${\cal S}_2$-orbit of $U$ minus
the average entanglement of ${\cal S}_2$ itself.
Notice  that the first two terms of Eq. (\ref{ep})
define two independent elements of the ring of polynomial
invariants of $U$ \cite{inv}.

Clearly the  simple relation \ref{ep})
holds just when $E$ is the linear entropy.
On the other hand the  structure of  \ref{ep}) ensures that $e_p$ is good entangling power measure
whatsoever a good entanglement measure $E$ is chosen.
It  is therefore tempting to suggest to use a non-decreasing real-valued smooth function of Eq. (\ref{ep})
vanishing at $0$   in order to {\em define} an entangling power measure for {\em any} $E.$

Employing  Eq.  (\ref{ep}) it is very easy to get
the  upper bound on the entangling power derived in \cite{ZAZA} (Eq. (9) with $d_1=d_2=d$).  
Since $E(S)=\mbox{max}_U\, E(U)= (d^2-1)/d^2$ one from Eq. (\ref{ep}) immediately obtains
\begin{equation}
e_p(U)\le   N_d^{-2}\, E(S)= \frac{d-1}{d+1}
\label{bound}
\end{equation}
Furthermore it is clear that in order for a $U$ to meet such bound i.e, to be {\em optimal}
in the language of Ref. \cite{ZAZA},  
a unitary $U$ must satisfy the constraints
\begin{equation}
E(U)=E(U\,S)=E(S).
\label{constra}
\end{equation}
Then an optimal transformation $U$ must be maximally entangled,
and such that  $U\,S$  is maximally entangled as well.
In terms of the ${\cal S}_2$-action discussed above
one can state the  optimal transformations
are characterized by the fact that the entanglement of $U$  is constant  along
its ${\cal S}_2$-orbit and maximal.
In view of its  simplicity  such a statement might be helpful in the search,
for optimal unitaries \cite{note}.
Notice  that  property (\ref{constra}) is obviously maximally violated by
unitaries belonging to the local equivalence classes of the identity and of  the swap. 
 
In this paper we discussed the notion of entanglement of a quantum evolution.
This has been done by simply observing that  all the notions developed for quantum-state entanglement
make sense for quantum evolutions   as well. Indeed operators acting on multi-partite
quantum state-spaces belong on their own to multi-partite Hilbert spaces.
This allows one to introduce entanglement measures $E$ for unitary transformations and then
to extend them to general quantum operations.
Adopting as  $E$ the  linear entropy one can obtain  analytical expressions for the  entanglement
of a unitary $U$ 
and   make a simple  connection  with its  entangling power.
Here we focused on the bi-partite case but it  should be clear
that the   main idea of lifting the notion of entanglement to the operatorial level
extends in a straightforward 
manner to the multi-partite case \cite{multien}.

I would like to thank Ch. Zalka for critical  discussions
and M. Rasetti for a careful  reading of the manuscript.

\end{multicols}


\begin{references}
\bibitem{expl} 
 Eq.  (\ref{op-ent}) can be written as
 $\overline{E(U)}^U=1-d^{-4} < T_{13}|\,\pi\,|T_{13}>$
where
$
\pi:=\int d U\,  |U><U|^{\otimes\,2},
$ is the   projector over the subspace of ${\cal H}_{HS}^{\otimes\,2}\otimes{\cal H}_{HS}^{\otimes\,2}$
commuting with all the $U^{\otimes\,2}$'s \cite{SYM}.
Such  subspace is spanned  by $\openone$
and  $T_{13}\,T_{24}$ \cite{CORN}.
An 
orthonormal basis is given by
$|0> :=\openone/d^2,\,
|1>:= (d^4-1)^{-1/2}\,(T_{13}\,T_{24}- \openone/d^2)$
Using this basis one has
$\pi=\sum_{i=0,1}|i><i|, $ hence
$ <T_{13}\,|\pi\,| T_{13}>=\sum_{i=1,2} |<i| T_{13}>|^2,$
from which  Eq. (\ref{ave}) follows.
\end{references}
\end{document}